\documentclass[]{iopart}
\usepackage{iopams}
\usepackage{setstack}
\usepackage{graphicx}
\usepackage{color}
\usepackage[english]{babel}
\usepackage{cite}
\usepackage{ulem}
\usepackage{upgreek}

\newcommand{\Si}{^{28}{\rm Si\,ball}}
\newcommand{\K}{{\rm K}}
\newcommand{\X}{{\rm X}}
\newcommand{\Y}{{\rm Y}}
\newcommand{\g}{{\rm g}}
\newcommand{\ii}{{\rm i}}
\newcommand{\ff}{{\rm ff}}

\newcommand{\ds}{\displaystyle}
\usepackage[colorinlistoftodos]{todonotes}

\begin{document}

\title[The kilogram: inertial or gravitational mass?]{The kilogram: inertial or gravitational mass?}{}%
\author{G Mana$^{1,2}$ and S Schlamminger$^3$}%
\address{$^1$INRIM -- Istituto Nazionale di Ricerca Metrologica, Torino, Italy}
\address{$^2$UNITO -- Universit\`a di Torino, Dipartimento di Fisica, Torino, Italy}
\address{$^3$NIST -- National Institute of Standard and Technology, 100 Bureau Drive, Gaithersburg, MD 20899, USA}
\ead{stephan.schlamminger@nist.gov}

\begin{abstract}
With the redefinition of the international system of units, the value of the Planck constant was fixed, similarly to the values of the unperturbed ground state hyperfine transition frequency of the $^{133}$Cs atom, speed of light in vacuum. Theoretically and differently from the past, the kilogram is now explicitly defined as the unit of inertial mass. Experimentally, the kilogram is realized by atom count or the Kibble balance. We show that only the former method measures the inertial mass without assuming the universality of free fall. Therefore, the agreement between the two measures can be interpreted as a test of the equivalence principle.
\end{abstract}

\submitto{Metrologia}

\maketitle
\ioptwocol

\section{Introduction}
A constant that is never made explicit links inertial to gravitational mass of all matter and energy. The inertial mass, $m_\ii$, determines the force required to accelerate an object by a given rate, $m_\ii = F/a$. The gravitational mass, $m_\g$, determines the force in Newton’s law of universal gravitation, $F = G m_\g M_\g\big/r^2$ and plays a role similar to the charge in Coulomb’s law.

Since Newton's unification of the Earth and celestial mechanics, the equivalence principle states that they are the same quantity. It implies the universality of free fall: in a gravitational field, locally, all bodies fall with the same acceleration, independently of their composition. Accepting this principle, the constant linking the gravitational and inertial masses must be dimensionless and can be conveniently set equal to one.

 The equivalence principle is an axiom of physics. Historically, the embrace of its validity led to the development of Newton's theory of gravity~\cite{Newton1687}, as well as of its successor, General Relativity~\cite{Einstein1915}. Upper limits to the principle violation must be established experimentally. Suppose that, for a material X, the ratio of gravitational to inertial masses is $m_\g(X)/m_\ii(X)=1+\eta(\X)$. Then, by comparing the free-fall acceleration of two different materials $X$ and $Y$, the statement
\begin{equation}
 \frac{a(\X)}{a(\Y)} = \frac{\ds 1+\eta(\X)}{\ds 1+\eta(\Y)} \approx 1 + \eta(\X) - \eta(\Y)
\end{equation}
can be made on the difference of the $\eta$ values.

The experimental tests date back to Galilei~\cite{Galilei1638} and potentially to earlier. A large sensitivity advance was made when E\"otv\"os~\cite{Eotvoes1922} realized that the sought difference, instead of being calculated from two large measurement results, $a(\X)$ and $a(\Y)$, as Galilei did, can be experimentally obtained via a null experiment, for example,  by using a torsion balance.  Such an experiment produces null if $\Delta \eta(\X,\Y):=\eta(\X) -\eta(\Y)$ is smaller than its sensitivity. Since then, large reductions of the $\Delta \eta$ upper limit have been reported~\cite{Roll1964,Braginsky1972,Schlamminger2008, Touboul2017}.

It is worth noting that all tests involve at least two materials, here symbolised by X and Y, and deliver only a statement about  an upper limit of the specific $\Delta\eta(\X,\Y)$

In 1901, the 3rd Conférence Générale des Poids et Mesures declared that the unit of mass {\it is equal to the mass of the international prototype of the kilogram}, a Pt-Ir artefact. This definition, consistent with the equivalence principle, did not distinguish between the inertial and gravitational masses. However, disseminating the kilogram by balances, we compared gravitational masses. In this sense, the quantity that was traced back to the Pt-Ir prototype was the gravitational mass.

The 2019 redefinition of the international system of units changed this state of affairs \cite{Wiersma_2021}. The unit of mass is now traced back to the stipulated values of the unperturbed ground state hyperfine transition frequency of the $^{133}$Cs atom, $\nu_{\rm Cs}$ speed of light in vacuum, $c$, and Planck constant, $h$. Therefore, the mass defect between the two hyperfine ground-state levels of the $^{133}$Cs atom is exactly $\Delta m_{\rm Cs} = h\nu_{\rm Cs}\big/c^2$. The roots of $\Delta m_{\rm Cs}$ are the Einstein and Planck equations $E=mc^2$ and $E=h\nu$. In special relativity, the equality of energy and the inertial (rest) mass follows from the conservation law for the energy-momentum tensor \cite{Ohanian_2012}, where gravity plays no role. Hence, $\Delta m_{\rm Cs}$ relates the inertial masses of the $^{133}$Cs atom before and after the transition, and the kilogram is now the unit of the inertial mass.

The object of this short communication is to examine how the equivalence principle underlies the practical realizations of the kilogram via counting atoms and the Kibble balance.

\section{Atom count}
We conceptually describe how the atom counting method can be used to realize an inertial mass standard. The first realisation step is recoiling $^{133}$Cs or $^{87}$Rb atoms by photons in an atom interferometer to measure the ratios between their inertial masses and the Planck constant \cite{Clade_2016,Yu_2019}. Alternatively, one can derive the $m_e/h$ ratio from the measured value of the Rydberg constant via hydrogen spectroscopy. The ratio comes into the Rydberg constant from the kinetic term of the hydrogen-atom Hamiltonian; therefore, $m_e$ stands for the inertial mass of the electron.  These mass ratios fix the absolute scale of atomic (inertial) masses via relative mass spectrometry by Penning traps.

In the second step, the kilogram is realised by atom counting. To determine the count in practice, a $^{28}$Si monocrystal is shaped as a quasi-perfect ball; the number $N_{\rm Si}$ of atoms in it is obtained from the measurement of the ball volume $V$ and lattice parameter $a_0$ according to $8V\big/a_0^3$, where $a_0^3\big/8$ is the atom volume, and 8 is the number of atoms in the cubic unit cell.

Making reference, for instance, to the $m_e/h$ quotient, the measurement equation is
\begin{equation}\label{mass}
\frac{m_\ii(\Si)}{h} = \frac{8V}{a_0^3}\frac{M(^{28}{\rm Si})}{M(e)}\frac{m_e}{h} ,
\end{equation}
where $m_\ii(\Si)$ is the ball's inertial mass and $M({\rm X})$ indicates the X's molar mass \cite{Massa:2020}. Since Si crystals are never perfect, mono-isotopic, and pure, (\ref{mass}) is corrected for the isotope abundances, impurities, and point defects (vacancies and interstitials). Also, the ball surface is characterised to correct for the oxide layer, adsorbed or absorbed water, and contaminants. In principle, one should take the mass defect associated with the binding energy of the atoms into account, but this correction is negligible at the present level of accuracy.

\section{Kibble balance}
Tracing mass measurements back to the Planck constant by a Kibble balance does not exactly imply the realisation of an inertial mass. Conceptually, a Kibble balance compares the power $m_\g(\K) gv$ generated by a gravitational mass $m_\g(\K)$ falling with constant velocity $v$ in a locally uniform gravitational field $g$, with the power $\mathcal{E}I$ dissipated by magnet-coil brake that would keep the mass motion uniform ($\mathcal{E}$ and $I$ are the electromotive force and eddy current).

In practice, the balance's measurement-equation,
\begin{equation}\label{Kibble}
 m_\g(\K) gv = \mathcal{E}I ,
\end{equation}
is assembled in two steps. Firstly, one measures the current $I$ necessary to hold up the mass in the brake's magnetic field. Next, the electromotive force $\mathcal{E}$ is measured at the ends of the brake coil when the mass moves with constant velocity $v$. The tie to the Planck constant is provided by two electrical quantum standards. $\mathcal{E}=n f/K_{\rm J}$ is measured in terms of the Josephson constant $K_{\rm J} = 2e/h$, where $n$ is an integer, $e$ is the elementary charge, and $f$ is a frequency \cite{Schlamminger:2019}. The current $I$ is converted into a voltage that is again measured against $K_{\rm J}$, passing it through a resistor that is a know fraction of the von-Klitzing constant $R_{\rm K} = h/e^2$.

The local gravitational field is determined by tracking a free-falling body, a corner-cube mirror, with a laser interferometer. Let's assume that $m_\ii(\ff)$ and $m_\g(\ff)$ are the inertial and gravitational masses of the free falling body. It is
\begin{equation}
 m_\g(\ff) g = m_\ii(\ff) a,
\end{equation}
where $a$ is the kinematic acceleration observed by measuring the traveled distance, $z(t)= z_0 + v_o t + a t^2/2$. A curve fitting procedure yields $a$, and $g$ is obtained as
\begin{equation}\label{g}
 g = \frac{m_\ii(\ff)}{m_\g(\ff)} a.
\end{equation}

Alternatives to the widely used classical gravimeters are measurements using atom or neutron interferometry and Bloch's oscillations of cold atoms in an optical lattice, which employ freely falling neutrons or atoms \cite{Colella_1975,Kasevich_1992,Clade_2005}.

A careful analysis of neutron interferometers by Littrell and coworkers \cite{Littrell_1997} shows that the difference of the quantum-mechanical phase accumulated by neutrons travelling the interferometer, $\Delta\Phi$, scales like $g m_\g(n)$. However, this phase difference is measured as a fraction of the de Broglie wavelength, $\lambda = h/p$, with $p = m_\ii(n)v$ and $v$ the neutrons' velocity. Thus,
\begin{equation}\label{Phi}
 \Delta\Phi \propto \frac{m_\g(n)}{m_\ii(n)} g
\end{equation}

Similar reasoning can be applied to atom interferometers and Bloch's oscillations in an optical lattice. The only difference is that the observable is not the de Broglie wavelength but the velocity and momentum changes of the free-falling atoms as probed by photon absorption \cite{Stuhler_2003,Clade_2005,Wolf_2011}. This process is kinematical and, hence, only sensitive to the inertial mass $m_\ii$.

Eventually, regardless of whether the probe mass is a macroscopic body, a neutron, or an atom, the measured value of $g$ is essentially given by (\ref{g}).
Therefore, by using (\ref{g}) in (\ref{Kibble}), the mass value that is obtained by using a Kibble balance is
\begin{equation}\label{final}
 m_\mathrm{KB}(\K) :=m_\g(\K) \frac{m_\ii(\ff)}{m_\g(\ff)} = \frac{\mathcal{E}I}{av} ,
\end{equation}
or, using the $\eta$ symbol introduced earlier,
\begin{equation} \label{mreport}
 m_\mathrm{KB}(\K) :=\frac{\ds m_\g(\K)}{\ds 1+\eta(\ff)} =  \frac{\ds 1+\eta(\K)}{\ds 1+\eta(\ff)} m_\ii(\K) .
\end{equation}
If the equivalence principle is assumed to hold true for the dropping object but not for the weighed one, i.e., $\eta(\ff)=0$ and $\eta(\K)\ne 0$, then the Kibble balance gives the gravitational mass $m_\g(\K)$. Contrary, if $\eta(\ff)=\eta(\K)$, which could be achieved by dropping $\K$, i.e., $\ff=\K$, then it gives the inertial mass $m_\ii(\K)$.  Lastly, for the case $\eta(\ff)=\eta(\K)=0$, there is no distinction between the two types of masses.

A suggestion to eliminate the weighing dependence on gravity by operating the Kibble balance horizontally via mechanical or electrostatic suspensions and measuring the inertial acceleration $a$ of the test mass was made by Cabiati \cite{Cabiati_1991} and further investigated by Kibble and Robinson \cite{Kibble_2014}. In this arrangement, the inertial force $m_\ii(\K)a$ substitutes for the gravitational force $m_\g(\K)g$ in (\ref{Kibble}), and inertial mass is measured without reference to the equivalence principle. As of today, Cabiati-type balances do not play a role in mass metrology and are not further discussed.

\section{Conclusions}
Contrary to the past, when balances disseminated gravitational masses, the kilogram is now per definition via $h$ the unit of inertial mass. However, only the atom count determines the inertial mass of the kilogram realisation without any reference to the equivalence principle.

Conceptually, as discussed in the previous section, the quantity measured by the Kibble balance depends on the assumptions made on the equivalence principle. Recent experiments \cite{Adelberger:2009,Wagner_2012} have shown the equivalence principle to hold at an uncertainty level that is many orders of magnitude smaller than the relative uncertainty obtained by Kibble balances. Hence, for all practical purposes, the question about what mass is measured is irrelevant.

Let us suppose that the mass of the same $\Si$ is measured by both counting the atoms and the Kibble balance and that the mass values are found in agreement. The first key comparison of the kilogram realizations based on the fixed numerical value of the Planck constant \cite{Stock_2020,Davidson_2021} is an embodiment of the supposed experiment, albeit in more than one step. The outcome is $m_\mathrm{KB}(\Si) =m_\ii(\Si)$. Combining this identity with (\ref{mreport}), where $\Si$ must substitute for $\K$, yields
\begin{equation}
 \eta(\ff) -\eta(\Si) = 0
\end{equation}
and an upper limit of $\Delta\eta(\ff, \Si)$ is obtained. Just as it was for Galileo's experiment, two large measurement results have been compared. It is, hence, not surprising that the
sensitivity, approaching 10 $\upmu$g/kg at the best \cite{Davidson_2021}, is not competitive against that of null tests \cite{Adelberger:2009,Wagner_2012}, whose relative sensitivities reach $10^{-13}$.

Comparing different kilogram realizations will never result in a competitive test of the equivalence principle. However, the musings in this article should not be dismissed for two  special features of that comparison. Firstly, in contrast to null tests, the result reported here critically depends on the absolute weighing of a 1 kg body. Secondly, unlike Eötvös-like experiments, which compare gravitational and inertial accelerations, electromagnetic and gravitational accelerations are compared here.

\section*{Acknowledgments}
Both authors want to thank the anonymous reviewers for their comments that helped clarify the text. G~M received support  from the Ministero dell'Istruzione, dell'Università e della Ricerca.

\section*{References}
\bibliography{equivalence_0523}   % name your BibTeX data base

\providecommand{\newblock}{}
\begin{thebibliography}{10}
\expandafter\ifx\csname url\endcsname\relax
  \def\url#1{{\tt #1}}\fi
\expandafter\ifx\csname urlprefix\endcsname\relax\def\urlprefix{URL }\fi
\providecommand{\eprint}[2][]{\url{#2}}
% Bibliography created with iopart-num v2.1
% /biblio/bibtex/contrib/iopart-num

\bibitem{Newton1687}
Newton S~I 1687 {\em {N}ewton's Principia - {T}he Mathematical Principles of
  Natural Philosophy (englisch translation from 1846)\/} (New York, 45 Liberty
  Street: D. Adee)

\bibitem{Einstein1915}
Einstein A 1915 {\em Sitzungsberichte der K{\"o}niglich Preu{\ss}ischen
  Akademie der Wissenschaften (Berlin)\/}  844--847

\bibitem{Galilei1638}
Galilei G 1638 {\em {D}icorsi e dimostrazioni matematiche intorno a due nuove
  scienze\/} (Leida: D. Adee)

\bibitem{Eotvoes1922}
v~E\"otv\"os R, Pek\'{a}r D and Fekete E 1922 {\em Ann. Phys.\/} {\bf 68}
  11--66

\bibitem{Roll1964}
Roll P, Krotkov R and Dicke R 1964 {\em Annals of Physics\/} {\bf 26} 442--517
  ISSN 0003-4916

\bibitem{Braginsky1972}
Braginsky V~G and Panov V~I 1972 {\em Soviet Physics JETP\/} {\bf 34} 463--467

\bibitem{Schlamminger2008}
Schlamminger S, Choi K~Y, Wagner T~A, Gundlach J~H and Adelberger E~G 2008 {\em
  Phys. Rev. Lett.\/} {\bf 100}(4) 041101

\bibitem{Touboul2017}
Touboul P {\em et~al.\/} 2017 {\em Phys. Rev. Lett.\/} {\bf 119}(23) 231101

\bibitem{Wiersma_2021}
Wiersma D~S and Mana G 2021 {\em Rendiconti Lincei. Scienze Fisiche e
  Naturali\/} {\bf 32} 655--663

\bibitem{Ohanian_2012}
Ohanian H~C 2012 {\em American Journal of Physics\/} {\bf 80} 1067--1072

\bibitem{Clade_2016}
Clad{\'{e}} P, Biraben F, Julien L, Nez F and Guellati-Khelifa S 2016 {\em
  Metrologia\/} {\bf 53} A75--A82

\bibitem{Yu_2019}
Yu C, Zhong W, Estey B, Kwan J, Parker R~H and Müller H 2019 {\em Annalen der
  Physik\/} {\bf 531} 1800346

\bibitem{Massa:2020}
Massa E, Sasso C~P and Mana G 2020 {\em MAPAN\/} {\bf 35} 511--519

\bibitem{Schlamminger:2019}
Schlamminger S and Haddad D 2019 {\em Comptes Rendus Physique\/} {\bf 20}
  55--63

\bibitem{Colella_1975}
Colella R, Overhauser A~W and Werner S~A 1975 {\em Phys. Rev. Lett.\/} {\bf
  34}(23) 1472--1474

\bibitem{Kasevich_1992}
Kasevich M and Chu S 1992 {\em Applied Physics B\/} {\bf 54} 321--332

\bibitem{Clade_2005}
Clad{\'{e}} P, Guellati-Kh{\'{e}}lifa S, Schwob C, Nez F, Julien L and Biraben
  F 2005 {\em Europhysics Letters ({EPL})\/} {\bf 71} 730--736

\bibitem{Littrell_1997}
Littrell K~C, Allman B~E and Werner S~A 1997 {\em Phys. Rev. A\/} {\bf 56}(3)
  1767--1780

\bibitem{Stuhler_2003}
Stuhler J, Fattori M, Petelski T and Tino G~M 2003 {\em Journal of Optics B:
  Quantum and Semiclassical Optics\/} {\bf 5} S75--S81

\bibitem{Wolf_2011}
Wolf P, Blanchet L, Bord{\'{e}} C~J, Reynaud S, Salomon C and Cohen-Tannoudji C
  2011 {\em Classical and Quantum Gravity\/} {\bf 28} 145017

\bibitem{Cabiati_1991}
Cabiati F 1991 {\em IEEE Transactions on Instrumentation and Measurement\/}
  {\bf 40} 110--114

\bibitem{Kibble_2014}
Kibble B~P and Robinson I~A 2014 {\em Metrologia\/} {\bf 51} S132--S139

\bibitem{Adelberger:2009}
Adelberger E, Gundlach J, Heckel B, Hoedl S and Schlamminger S 2009 {\em
  Progress in Particle and Nuclear Physics\/} {\bf 62} 102--134

\bibitem{Wagner_2012}
Wagner T~A, Schlamminger S, Gundlach J~H and Adelberger E~G 2012 {\em Classical
  and Quantum Gravity\/} {\bf 29} 184002

\bibitem{Stock_2020}
Stock M, Concei{\c{c}}{\~{a}}o P, Fang H, Bielsa F, Kiss A, Nielsen L, Kim D,
  Kim M, Lee K~C, Lee S, Seo M, Woo B~C, Li Z, Wang J, Bai Y, Xu J, Wu D, Lu Y,
  Zhang Z, He Q, Haddad D, Schlamminger S, Newell D, Mulhern E, Abbott P,
  Kubarych Z, Kuramoto N, Mizushima S, Zhang L, Fujita K, Davidson S, Green
  R~G, Liard J~O, Murnaghan N~F, Sanchez C~A, Wood B~M, Bettin H, Borys M,
  Mecke M, Nicolaus A, Peter A, Müller M, Scholz F and Schofeld A 2020 {\em
  Metrologia\/} {\bf 57} 07030

\bibitem{Davidson_2021}
Davidson S and Stock M 2021 {\em Metrologia\/} {\bf 58} 033002

\end{thebibliography}

\end{document}